\documentclass{article}

\usepackage{arxiv}

\usepackage[utf8]{inputenc}
\usepackage[T1]{fontenc}
\usepackage{hyperref}
\usepackage{url}
\usepackage{booktabs}
\usepackage{amsfonts}
\usepackage{nicefrac}
\usepackage{microtype}
\usepackage{graphicx}
\usepackage{natbib}
\usepackage{doi}

\title{Addressing the Synergy Gap:\\ The Six Elements of the Design Space}

\author{
  \href{https://orcid.org/0000-0001-6826-9688}{\includegraphics[scale=0.06]{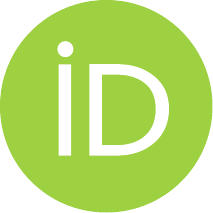}\hspace{1mm}Tommaso Turchi}\thanks{Corresponding author.} \\
  Department of Computer Science \\
  University of Pisa \\
  Pisa, Italy \\
  \texttt{tommaso.turchi@unipi.it} \\
  \And
  \href{https://orcid.org/0009-0004-5663-5854}{\includegraphics[scale=0.06]{orcid.pdf}\hspace{1mm}Ben Wilson} \\
  Computational Foundry \\
  Swansea University \\
  Swansea, United Kingdom \\
  \texttt{b.j.m.wilson@swansea.ac.uk} \\
  \And
  \href{https://orcid.org/0000-0002-1486-5537}{\includegraphics[scale=0.06]{orcid.pdf}\hspace{1mm}Matt Roach} \\
  Computational Foundry \\
  Swansea University \\
  Swansea, United Kingdom \\
  \texttt{m.j.roach@swansea.ac.uk} \\
  \And
  \href{https://orcid.org/0000-0002-5242-7693}{\includegraphics[scale=0.06]{orcid.pdf}\hspace{1mm}Alan Dix} \\
  Cardiff Metropolitan University, United Kingdom \\
  \& Computational Foundry, Swansea University \\
  \texttt{alan@hcibook.com} \\
  \And
  \href{https://orcid.org/0000-0002-2601-7009}{\includegraphics[scale=0.06]{orcid.pdf}\hspace{1mm}Alessio Malizia} \\
  Department of Computer Science, University of Pisa, Italy \\
  \& Faculty of Logistics, Molde University College, Norway \\
  \texttt{alessio.malizia@unipi.it} \\
}

\renewcommand{\shorttitle}{Addressing the Synergy Gap: The Six Elements of the Design Space}

\hypersetup{
  pdftitle={Six Elements for Human-AI Decision Systems},
  pdfsubject={cs.HC, cs.AI},
  pdfauthor={Tommaso Turchi, Ben Wilson, Matt Roach, Alan Dix, Alessio Malizia},
  pdfkeywords={synergistic interaction, decision-making, design, explainable AI, artificial intelligence},
}

\begin{document}
\maketitle

\begin{abstract}
AI is now embedded in healthcare, finance, policy, and many other domains, yet genuine human-AI synergy --- combined performance that exceeds what either party achieves alone --- is uncommon. Meta-analyses show that AI assistance tends to improve human performance compared to working alone, but studies finding true synergy are scarce. We call this persistent shortfall the synergy gap. Most current work treats human-AI combination as an engineering problem and concentrates on interpretability, trust calibration, or interface design. These matter, but they cover only part of what determines whether combination works. Closing the synergy gap, we argue, requires explicit engagement with a wider design space. We map that space through six interconnected elements: sociotechnical context, decision-making frameworks, human decision participants, AI capabilities, interaction, and holistic evaluation. For each element, we describe what it covers, how it shapes the others in practice, and what it implies for design. The result is a shared vocabulary for practitioners building hybrid systems, an analytical lens for researchers studying combination patterns, and a starting point for evaluators interested in the full quality of human-AI decision-making rather than accuracy alone.
\end{abstract}


\section{Introduction}

Imagine a radiologist working with an AI system that doesn't just flag potential tumours, but engages in a diagnostic conversation --- questioning assumptions and suggesting alternative interpretations in real-time. Or consider an urban planner whose AI ``partner'' helps navigate complex stakeholder priorities, not by optimising a single metric, but by facilitating nuanced trade-offs between environmental impact, community needs, and economic constraints.

This vision of human-AI synergy feels tantalisingly close, yet remains largely unrealised. While AI systems increasingly support decision-making across healthcare, finance, policy, and beyond, most still operate within narrow combination paradigms: humans validate, override, or defer to AI recommendations. The result is what we call the ``synergy paradox'' --- while some systems improve individual performance, there is a gap, rarely bridged, between the observed outcome and the ideal potential where human-AI combinations exceed what either could accomplish alone.

This challenge inspired the SYNERGY Workshop series \citep{synergy2024,synergy2025}, held for its second edition in Pisa as part of the broader TANGO European Project. Rather than proposing yet another algorithmic solution, we sought to map the design space itself: What elements do designers need to create genuinely synergistic human-AI decision systems? How do these elements interact across different domains, contexts and stakeholder groups? What has the community learned from attempts to bridge the synergy gap?

This paper describes the design space that emerged from this collaborative inquiry as a shared understanding for thinking about building effective systems for human-AI combination in decision-making contexts. Drawing on workshop discussions, participant feedback, and analysis of contemporary research, we present a set of six interconnected elements that shape how humans and AI should work together effectively to yield better decisions.

\section{The Synergy Gap}

The promise of human-AI synergy traces back to Joseph Licklider's 1960 vision of ``man-computer symbiosis'' --- a future where computers would augment human intellect rather than simply automate tasks \citep{licklider1960ManComputerSymbiosis}. Today, we see glimpses of this vision: AI systems help doctors diagnose diseases, assist analysts in making sense of complex data, and support planners in optimising resource allocation.

\begin{figure}[!ht]
    \centering
    \includegraphics[width=0.7\linewidth]{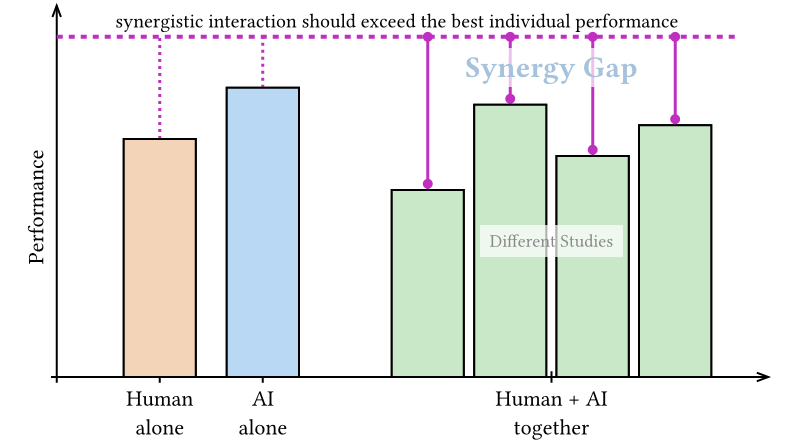}
    \caption{The synergy gap (exemplified) in human-AI decision-making. Different studies show varying degrees of this gap, highlighting the persistent challenge of achieving genuine human-AI synergy.}
    \label{fig:synergy}
\end{figure}

Yet research reveals a persistent gap between potential and practice. While meta-analyses \citep{vaccaroWhenCombinationsHumans2024,
laiScienceHumanAIDecision2023} show that AI assistance typically improves human performance compared to working alone, studies demonstrating true synergy --- where human-AI collaboration exceeds the performance of the best individual agent --- remain rare \citep{jacobsHowMachinelearningRecommendations2021,vaccaroWhenCombinationsHumans2024,bergerFosteringHumanLearning2025}. This synergy gap reflects deeper challenges in designing and deploying AI systems, as depicted in Figure \ref{fig:synergy}.

To be effective, combination should close the synergy gap. Current approaches often treat human-AI combination as an engineering problem: How can we make AI outputs more interpretable? How do we calibrate user trust? While these questions matter, they miss a crucial insight: effective human-AI combination emerges from carefully orchestrating multiple interacting elements, not just better algorithms or interfaces \citep{hemmerComplementarityHumanAICollaboration2024,laiScienceHumanAIDecision2023}.

\section{Elements of Human-AI Decision System Design}
\begin{figure}[!ht]
    \centering
    \includegraphics[width=\linewidth,trim={0 0 4cm 0},clip]{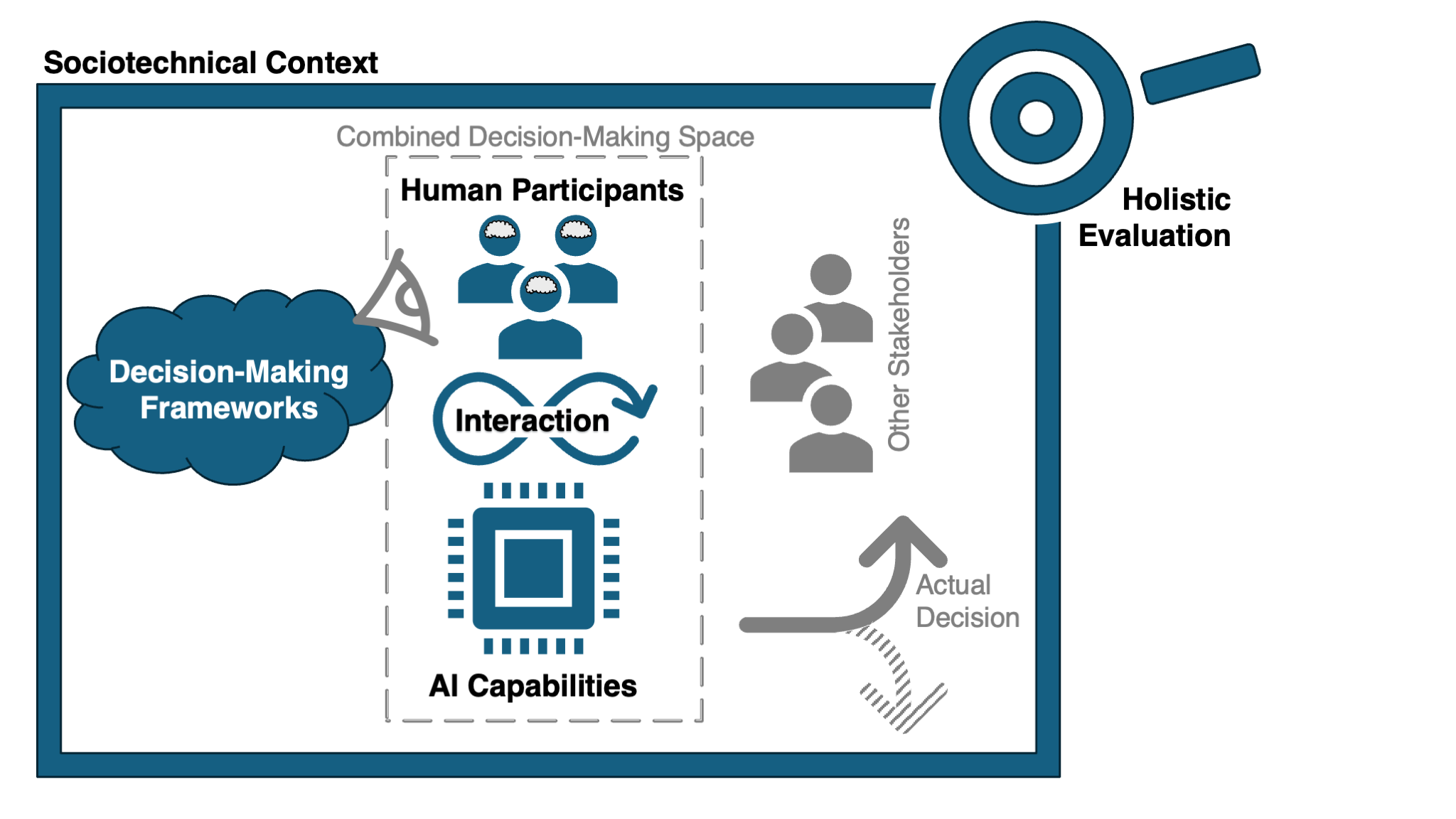}
    \caption{The six elements of the design space for human-AI decision systems.}
    \label{fig:sixelements}
\end{figure}

Our design space comprises six key elements shaping effective human-AI combinations in decision-making contexts (Figure \ref{fig:sixelements}). The form of expression of each component may differ depending on the decision context itself.

\subsection{Sociotechnical Context}
The sociotechnical context --- which includes the goal and the conditions under which it is addressed --- fundamentally shapes what effective human-AI combination looks like. This encompasses not just the technical constraints of the domain, but the social, organisational, and cultural factors that influence how decisions are met, made and evaluated \citep{klein1993DecisionMakingAction}.

It is crucial to think about what constitutes a good solution, that is, the extent to which it achieves the \textbf{goal} of the overall system, \textbf{how clearly the decision problem can be specified} and whether stakeholders (decision-makers and decision-subjects) agree to this. In domains like chess or weather forecasting, problems are relatively well-defined, while in contexts like urban planning or allocating healthcare resources, problem definition becomes contested and politically charged. This profoundly influences what kinds of AI support are possible and appropriate.

The \textbf{social organisation of decision-making} also matters tremendously. To understand real-world deployments, it is crucial to move beyond individual ``users'' and consider multiple stakeholders \citep{puranamHumanAICollaborative2020,baoLiteratureReviewHumanAI2023} with different roles, expertise, exposure and authority levels, including those affected by decisions but with limited agency in the process. We need to ask: Who are the stakeholders? How do they relate to one another, the decision at hand, and the AI system? What control relations exist? What commitments are in place? Understanding these participation structures helps designers anticipate how AI systems will be received and integrated into existing dynamics.

\textbf{Contextual dynamics} capture another crucial dimension --- the rate and nature of change in the decision environment. Navigational scenarios require rapid adaptation to changing conditions, while social policy processes operate in relatively stable environments. Dynamic contexts often require more flexible interaction patterns and adaptive AI capabilities. Related to this is goal flexibility: whether objectives are stable and well-defined or shifting and contested. In optimisation problems, goals may be clear and unchanging, while in policy contexts, goals often evolve as stakeholders learn and negotiate.

\textbf{Time sensitivity} affects everything from interface design to the level of exploration or explanation that's feasible. Some decisions allow for deliberate consideration and extensive human-AI dialogue, while others require immediate response with minimal interaction overhead. Similarly, information certainty ranges from situations with clear, complete, reliable data to those with significant ambiguity, uncertainty, or missing information. This characteristic influences both the confidence in AI recommendations and the necessary level of human oversight.

Different contexts demand different levels of expertise and type of domain knowledge, from situations manageable by generalists to those requiring deep specialist expertise, often more than one person. This shapes not just what information to present, but how to present it and what level of explanation or justification is needed.

\textbf{Task orientation} is yet another aspect of the combination process. An AI system that augments human capability on a core task or on the majority of a human’s concurrent workload has greater leeway for requiring the human to concede to its operational needs. Whereas, a machine that provides assistance only on a marginal basis is likely to prove disruptive if its design isn’t carefully crafted to suit that role.

Finally, the \textbf{cardinality of participants} will inevitably affect what interactions they can have and to what effect. If we move beyond the common human-machine dyad, we can recognise multiple participants of each type \citep{salimzadehWhenDoubtUnderstanding2024,askarisichaniPredictiveModelsHuman2022}. This introduces both composite and complementary capabilities. However, it also introduces complexity in resolving different positions into a single outcome.

\subsection{Decision-Making Frameworks}
This often overlooked element captures the variety of underlying assumptions that will be present about how decisions should be made and how humans make decisions in combination with others and with machines \citep{bellDESCRIPTIVENORMATIVEPRESCRIPTIVE1988}. These assumptions, often implicit, profoundly shape system design and interaction patterns.

\textbf{Normative views} assume that people should make optimal decisions and design systems to support formal rationality. Utility and Bayesian decision theories exemplify this approach \citep{aminBayesianFrameworkHumanAI2026}, working well in domains with clear objectives and quantifiable trade-offs. AI systems designed from normative perspectives often focus on providing optimal recommendations or highlighting deviations from rational choice.

\textbf{Prescriptive views} take a more pragmatic stance, recognising human limitations while assuming people can be helped to make better decisions through appropriate support. Frameworks like the Cynefin model \citep{cynefin} or Vroom-Yetton-Jago \citep{vyj} decision trees guide when and how to provide different types of assistance. These approaches prove useful in semi-structured domains where some guidance improves outcomes without requiring complete optimisation.

\textbf{Descriptive views} focus on how people make decisions, embracing the reality that humans rely on heuristics, biases, and intuition rather than formal analysis. Kahneman's dual process theory \citep{kahnemanThinkingFastSlow2011}, naturalistic decision-making approaches \citep{klein1993DecisionMakingAction}, and similar frameworks recognise that human decision-making often works well despite --- or perhaps because of --- its departure from formal rationality. AI systems designed from descriptive perspectives usually focus on supporting rather than replacing human judgment processes.

Underlying these approaches are specific \textbf{theories about human cognition} that inform interaction design. Attention theories help determine how much information to present and when. Learning theories guide how systems should adapt to user behaviour over time. Memory and perception research influences information presentation formats. Situation awareness models help designers understand what information humans need to maintain effective oversight. Mental model theories guide how to help users understand AI capabilities and limitations. Theory of mind research informs how AI systems better understand and respond to human intentions and expectations.

The choice among these approaches isn't purely theoretical \citep{steyversThreeChallengesAIassisted2023} --- it has practical implications for everything from interface design to evaluation criteria. Systems based on normative views emphasise optimisation and efficiency, while those based on descriptive views prioritise supporting human sense-making and maintaining situational awareness. Often, we are unaware of our own theoretical assumptions. Designers must engage consciously with this element and reflect on its implications.

\subsection{Human Decision Participants}
Those directly involved in decision-making have a variety of characteristics.

\textbf{Expertise} can be distributed over multiple people: this expertise includes differing domain knowledge, skills, technical literacy, and system familiarity. Expertise might be enhanced through effective combination or degraded through over-reliance on automation \citep{inkpenAdvancingHumanAIComplementarity2022}. Understanding the dynamic of expertise evolution helps designers create systems that support rather than undermine human capability development.

\textbf{Cognitive characteristics} also shape how stakeholders can effectively engage with AI systems. Working memory capacity affects how much information can be processed simultaneously, while reasoning approaches influence how people prefer to approach problems. Attention control and executive function determine how well stakeholders can manage complex human-AI interactions. These aren't fixed constraints but rather capabilities that effective systems should support and complement.

We also consider the \textbf{psychological factors} that affect combined decision-making. Risk tolerance influences stakeholders' willingness to accept AI recommendations, while trust disposition affects initial reactions to AI systems. Uncertainty tolerance shapes how comfortable people are with ambiguous AI outputs, and emotional states like stress or fatigue can significantly impact the effectiveness of human-machine combination. Some of these are stable personality traits, while others are variable conditions that systems might need to detect and adapt to.

\textbf{Individual preferences} add another layer of complexity. People have different preferences regarding interaction pace, information granularity, and degree of control \citep{luMixMatchCharacterizing2024,farmerExaminingInteractionsUser2025}. For example, some prefer detailed explanations; some want concise summaries; some like maintaining tight control over the interaction; and some are comfortable with more autonomous AI behaviour. Effective systems need to accommodate this diversity while maintaining coherent synergistic processes.

\subsection{AI Capabilities}
There are many ways AI systems can serve decision contexts. Prediction capabilities allow systems to forecast future states or outcomes based on historical patterns. Classification helps organise and categorise complex information. Association and clustering capabilities can identify patterns and relationships that might not be obvious to human stakeholders. Optimisation functions can search for solutions within defined constraints. The most effective systems will combine multiple capabilities, acting as prediction engines in some contexts and pattern detectors in others.

There is, however, a question of \textbf{congruency} between the real problem and its formulation. In bringing AI capabilities together in a decision context, designers attempt to match the system’s decision space to the real-world problem space. We should always consider how well that match is achieved. 

There is another question that relates to the \textbf{adaptation} capabilities of the system. Does it learn from new interactions, adapt to user patterns, or remain static \citep{aminAlignWhenThey2026}? This affects how the human-AI combination evolves over time.

\textbf{Informational scope} covers whether the system can access the optimal source data to support this decision. What constraints are there on access? What effects may there be on the suitability of what is accessible and how it is available?

\textbf{Process transparency} involves the degree to which AI systems can make their reasoning accessible to human stakeholders. Some systems operate as black boxes where internal decision processes remain opaque, while others provide partial visibility into their reasoning processes. Transparency sometimes comes at the cost of performance or simplicity. The key challenge lies in presenting insights or explanations that minimise cognitive load while supporting appropriate reliance \citep{bansalDoesWholeExceed2020,bansalAccuracyRoleMental2019}.

Finally, \textbf{outcome transparency} is related to uncertainty handling. Can the system express confidence levels, handle ambiguous inputs, and flag when it's operating outside its training domain \citep{zhangEffectConfidenceExplanation2020,maWhoShouldTrust2023a}? Can it provide adjuvant (i.e., additional, supportive) output that it might not be capable of evaluating but that humans can use in the decision process \citep{guoExplainingImprovingInformation2025}? This transparency facilitates synergy.

\subsection{Interaction}
Interactions should be understood as adaptive relationships rather than fixed protocols. They may influence the fundamental AI model used as well as the design features.

One of the most fundamental considerations is how AI functions in relation to humans and its \textbf{role} in the combination process. AI systems might serve as peers, coaches providing guidance while preserving human agency, or provocateurs challenging assumptions and offering alternative perspectives \citep{sarkarAIShouldChallenge2024}. These roles needn't be fixed; they can shift dynamically based on context and human needs, much like how human team members might take on different roles in different situations.

We also consider the \textbf{mode of AI intervention} \citep{gomezHumanAICollaborationNot2023}. Some systems focus primarily on information presentation and insight generation, helping users understand complex data or identify patterns. Others emphasise critique and evaluation, analysing proposed decisions and highlighting potential issues. Still others provide specific recommendations or guidance. Understanding which of these purposes is most appropriate --- and when --- helps guide interaction design choices.

The distinct \textbf{nature of information obtainable} by humans and machines influences how they combine in decision-making. This can relate to how much overlap there is in the information each can access. But also, how much friction is encountered in obtaining new information that might be relevant to a decision.

The \textbf{content and format of information} exchange also shape human-AI combination significantly because it relates to the match between the system’s output (or input) space and the human’s input (or output) space. Some approaches work well with specific examples and precedents --- showing users similar cases or historical decisions to provide context. Others benefit from more abstract representations and general principles that help users understand underlying patterns or rules. Many effective systems combine both approaches, offering concrete examples when users need context and abstract models when they need to understand general principles.

\textbf{Control and initiative} represent another crucial design dimension. AI systems might be proactive, suggesting actions or raising concerns in anticipation, or reactive \citep{holterDeconstructingHumanAICollaboration2024}, responding to human queries and prompts. The most effective systems allow this balance to shift based on context and participant preferences. A system might be more proactive when detecting potential problems, but more reactive when users are in exploratory mode.

The \textbf{temporal organisation} of human-AI communication matters too. Some contexts demand continuous information flows. Others need to provide only sparse or intermittent intervention in response to conditions. Further contexts benefit from progressive disclosure that gradually reveals information to manage cognitive load. Others support trial-and-error approaches that encourage iterative exploration. Traditional turn-taking patterns work well in some domains, while others require more fluid, interleaved interaction \citep{vanberkelHumanAIInteractionIntermittent2021}.

The \textbf{modality of communication} --- whether through natural language, visual representations, or multimodal combinations --- should align with both the type of information being shared and user preferences.

\subsection{Holistic Evaluation}
The \textbf{value of decision outcomes} extends beyond simple accuracy measures \citep{turchiEcologicalValidityMissing2025,bucincaProxyTasksSubjective2020}. Sociotechnical fit --- how well the system aligns with organisational culture, human needs, and existing processes --- often determines long-term success. Stakeholder satisfaction with both the decision process and outcomes affects adoption and sustained use. Long-term sustainability encompasses not just technical maintenance but continued human engagement and skill development. There's growing recognition that we must consider impacts on wellbeing and human capability development, not just immediate decision quality.

Evaluation must go beyond traditional machine learning metrics to encompass the quality of human-AI combination and its broader impacts. This represents a significant shift from algorithm-centred thinking to combination-centred assessment, which accords with a goal-oriented evaluation perspective. Of course, this more holistic assessment should also drive the design process, effectively what it means to be a good design. 

Traditional quantitative metrics remain essential but need to be understood in context. Accuracy, consistency, fairness, and efficiency all matter, but raw performance numbers may matter less than the \textbf{quality of combination processes} in many contexts. Rather than optimising for AI accuracy alone, a system with slightly lower AI accuracy that helps humans understand and trust the AI might create a better overall decision outcome.

Process quality evaluation focuses on the decision-making journey itself. Systems should be evaluated not just on how each decision unfolds but also on whether the overall pattern of human-AI combination achieves the intended dynamics and maintains effectiveness across varied decision contexts over time. This includes asking whether systems preserve human autonomy and agency, rather than reducing people to rubber-stampers. 

Metrics such as appropriate reliance and trust calibration \citep{schemmerAppropriateRelianceAI2023,schoefferAIRelianceDecision2023} become particularly important: are stakeholders developing the right/appropriate confidence levels in AI recommendations?

Traditional usability principles around effectiveness, efficiency, and satisfaction remain relevant but must be applied to sustained combination rather than individual interactions. 

Integration with existing workflows is crucial --- systems that require complete process redesign face adoption challenges, while less disruptive solutions --- even if technically less advanced --- may succeed precisely because they fit more naturally into established practices.

We also highlight growing concerns about \textbf{provenance and accountability}. As decisions emerge from complex human-AI interactions, tracking who contributed what becomes challenging but essential. This is particularly important in contexts where decisions must be justified or where regulatory requirements demand clear accountability chains. Systems need to support not just good decisions but defensible and traceable decision processes \citep{ferrarioEpistemologyGivesFuture2026}.

\section{Implications for Design and Practice}
This design space offers several practical implications for those designing human-AI decision systems, though we emphasise that these suggestions emerge from our collaborative inquiry rather than prescriptive rules.

\paragraph{Build from Context, Not Technology} Before considering technical solutions, designers must understand the participation dynamics in their context: Who makes decisions? How do they relate to each other? What are the existing power dynamics and communication patterns? Environmental dynamics matter equally --- understanding whether the decision environment is stable or rapidly changing, whether goals are fixed or negotiable, and what time pressures exist. Information certainty and expertise requirements complete this contextual picture \citep{wilsonDimensionsHumanMachineCombination2025}. A system designed for stable, well-defined problems with clear expertise hierarchies will look very different from one supporting ambiguous, contested decisions among diverse humans involved.

Designers should recognise that cognitive theoretical perspectives inform their design work. They should consider how the perspectives they consciously or unconsciously employ influence the level of synergy they facilitate through their system design. Designers inevitably rely on assumptions about how people make decisions, whether through fast and intuitive judgments, deliberate analytical reasoning, or context-shaped heuristics. These perspectives function as conceptual lenses: they shape how designers see the problem space, what trade-offs they emphasise, and which design opportunities they pursue.

\paragraph{Combine Carefully} Congruence, adaptation, scope and transparency require that we approach combination with the real sociotechnical context in mind \citep{rastogiTaxonomyHumanML2022}. The forms of interaction that we facilitate need to serve the distinct contributions of humans and machines to the situated problem and allow them to fulfil their complementary potential. Of course, effective systems must be designed for human diversity rather than imagining a single, average user. Decision subjects, decision-makers, managers, domain experts, and affected communities have different needs, constraints, and evaluation criteria. This doesn't mean designing different systems for each group; instead, it means creating flexible systems that can adapt to support various human roles, relationships, and contexts.

\paragraph{Iterate and Evaluate Holistically} The most successful human-AI systems won’t be static tools but evolving partnerships that improve over time. Synergy can arise when users develop expertise with the system, learning its capabilities and limitations through repeated interaction \citep{bergerFosteringHumanLearning2025}. AI systems can improve through feedback, adapting to usage patterns and context-specific requirements. Meanwhile, organisational contexts inevitably shift to accommodate new possibilities --- workflows evolve, roles change, and new forms of combination emerge. Designers must anticipate and support these evolutionary processes rather than assuming their initial design will remain optimal over time. More than this, evaluation approaches must evolve beyond traditional metrics to assess the full quality of human-AI combination. This holistic evaluation perspective requires new methods and metrics that capture the richness of combination processes and their implications for the humans involved and affected. We should adopt overall combined system goals that express what it means to have synergy that improves human conditions.

\section{Conclusion}
Effective system design requires explicit awareness of the six elements. Awareness alone, however, is not enough. Designers must also enact these implications in their design choices.

The ideas presented here represent a collaborative effort to map the design space for human-AI decision systems. The six elements that constitute the design space are always present, but frequently ignored or overlooked in work on hybrid systems seen in the literature. This paper describes the whole design space: to achieve synergy in the complex interactions that shape human decision-making in combination with machines, we must engage with the entire space.

The workshop that informed this design space demonstrated the value of bringing together diverse perspectives from computer science, psychology, design, and domain expertise. The feedback we received emphasised the need to capture not just technical capabilities but the full sociotechnical context in which effective human-AI combinations occur.

Moving forward, bringing these six elements together can serve multiple purposes: as a design tool for practitioners building human-AI systems, an analytical lens for researchers studying combination patterns, and a foundation for more nuanced evaluation approaches that capture the full richness of human-AI synergies.

The goal is not to replace human judgment with AI optimisation, but to create systems where humans and AI can decide in combination --- each contributing their unique strengths to challenges that neither could address alone. Achieving this vision requires continued collaboration between technologists, domain experts, and the communities these systems aim to serve.

We invite others to engage with these ideas, test them in different contexts, and contribute to the ongoing conversation about designing AI systems that truly augment human capabilities rather than simply automating human tasks. The path to genuine human-AI synergy lies not in any technological breakthrough, but in the careful orchestration of the many elements that shape how humans and AI can work together effectively.

\section*{Acknowledgments}
This reflection emerged from the SYNERGY Workshop series and numerous discussions within the TANGO community. We thank all workshop participants for their insights and contributions to this ongoing dialogue about the future of human-AI combination.

Funded by the European Union. Views and opinions expressed are however those of the author(s) only and do not necessarily reflect those of the European Union or the European Health and Digital Executive Agency (HaDEA). Neither the European Union nor HaDEA can be held responsible for them.
Grant Agreement no. 101120763 - TANGO. 


\bibliographystyle{unsrtnat}
\bibliography{references} 

\end{document}